# Point Location in Constant Time

*Sairam Chaganti*

School of Science and Engineering
University of Missouri at Kansas City
Kansas City, MO 64110, USA
scmdt@umkc.edu

*Yijie Han*

School of Science and Engineering
University of Missouri at Kansas City
Kansas City, MO 64110, USA
hanyij@umkc.edu

## Abstract

We preprocess the input subdivision with n points on the plane in $O(n\sqrt{\log n})$ time to facilitate point location in constant time. Previously the preprocessing time is $O(n \log n)$ and point location takes $O(\log n)$ time.

*Keywords: sorting, subdivision, point location.*

# 1.Introduction

Point location is a problem in computational geometry that has been studied by many researchers [4, 5, 10, 11, 12]. The previous best results for this problem [4, 10, 11] take $O(n \log n)$ preprocessing time, $O(n)$ storage and $O(\log n)$ time for locating the point.

In this paper we show that we can preprocess the input subdivision in $O(n\sqrt{\log n})$ time to facilitate the point location in constant time.

The approach we take is to convert the real numbers of point coordinates to integers and then multiple integers can be packed into one word. This speeds up the algorithm.

The tool we used to do this real number to integer conversion is the real number sorting algorithm shown in [7] which has time $O(n\sqrt{\log n})$ time and runs on the computation model used in computational geometry.

# 2. The $O(n\sqrt{\log n})$ Time Sorting Algorithm for Sorting Real Numbers

It used to be that real numbers can only be sort with comparison sorting algorithms. Comparison sorting has a lower bound of $\Omega(n \log n)$ time complexity for sorting n numbers [3]. This is because for the input n numbers there are n! permutations of these n number and only one permutation (the right permutation) permutes the input n numbers to the sorted order assuming that these n input numbers are all different. A comparison of two input numbers a and b will have the result of a > b or a < b. Among the n! permutations of n input numbers some permutations A comply with the result of a >b and other permutations B comply with the result of a <b. At least one of A and B has no less than n!/2 permutation.

If a > b then the permutations in B need not be further considered as they do not satisfy a > b. But we need to further figure out which permutation in A is the right permutation. If a < b then permutations in A can be precluded as they do not satisfy a <b. But we need to figure out which permutation in B is the right permutation. In the worst case the comparison of a and b will preclude the smaller set of A and B and thus we have at least n!/2 permutations remaining. Each subsequent comparison will, in the worst case, cut the number of remaining permutations to half. Thus we need at least log(n!) comparisons to cut the number of permutations to 1. By Sterling's approximation n!≈(n/e)$^n$, where e is the natural number. Thus the number of comparisons needed is $\Omega(n \log n)$. This reasoning is taken from [13].

The way that we have achieved $O(n\sqrt{\log n})$ time for sorting real numbers is that we find a way to convert real numbers to integers such that the conversion process preserves the order of these input numbers. This conversion can be done in $O(n\sqrt{\log n}))$ time and linear space [7].

After converting to integers we can then sort these integers. Integer sorting takes $O(n \log \log n)$ time [8, 9].

## 3.Preprocess

We first convert the real numbers of the point coordinates to integers. This is accomplished by sorting the real numbers of x and y coordinates of n points. This is done in $O(n\sqrt{\log n})$ time [7] as mentioned above. We will do this separately for nonnegative numbers and negative numbers. Let $a_0$, $a_1$, …, $a_{2n-1}$ be sorted sequence of these coordinates. We separate negative numbers from positive numbers and do $a_{i+1}-a_i$ when $a_{i+1}$ and $a_i$ are both positive or both negative for i=0, 1, …, 2n-2, and find the most significant bit $b_i$ of $a_{i+1}-a_i$ (method given in the following). Here $b_i$ is counted as 0 for the least significant integral bit. If we cut $a_{i+1}$ and $a_i$ at bit $b_i$, that is: remove all bits less significant than the $b_i$-th bit, then the resulting numbers for $b_{i+1}$ and $b_i$ are integers (after shifting $–b_i$ bit or multiplying by $2^{-bi}$ ) and their order is preserved. We will find the smallest $b_i$ value b among the 2n-1 $b_i$ values and cut all $a_i$'s, 0 ≤ i < 2n, at bit b. This converts all real numbers of point coordinates to integers and the original order of these real numbers is preserved in the converted integers.

Real number a can be cut at bit b and convert to an integer by the operation int(a*2$^{-b}$).

The most significant bit can be found by the method shown in [6]. It can also be found by taking the logarithm base 2 then take the integer part of the result. Another approach is to first scale all real number so that their absolute value is less than 1. Then take the largest value b among $1/|a_{i+1} - a_i|$, i=0, 1, 2n-1. Value b can then be used to convert $a_i$ to an integer as

int($ba_i$). This converts all $a_i$'s to integers while preserving their order.

After converting real numbers for point coordinate values to integers we then triangulate the subdivision. This can be done in $O(n)$ time using [2].

We then pack the 2'complement values of the integer coordinate for x (y) coordinate into one word. Say that each integer has B bits, we will add an extra two bits of 00 (positive) or 01 (negative) in front of every integer value. The first bit is need to prevent overflow to the next integer when doing pairwise addition. The second bit is the sign bit. Thus we will use B+2 bits for each integer. Let the word for the packed x values be X and let the word for the packed y values be Y. For an edge with the equation $a_ix+b_iy+c_i=0$ we will also pack the converted integer values (of B+2 bits) of $a_i$, $b_i$, $c_i$ into one word for all edges i=0, 1, …, m-1.

We will also prepare some constants. The constants we need are $C_1=(0^{B+1}1)^n$, $C_2=(010^B)^n$. Each of these constant can be precomputed in $O(\log n)$ time. $C_1$ is used to duplicate a B bit integer a for n copies by $aC_1$ operation. $C_2$ is used to extract sign bits.

Example:
Line1: 1x+2y=0
Line2: 2x+4y-3=0
Line3: -5x+6y+7=0
.
.
.
.
.

We will evaluate point (10, 15) against these line equations.

Assuming that we use B=5 bits for each integer. In order for performing the multiplication for packed integers we use 2*(B+2)=14 bits for each integer. Because for multiplication the signs are multiplied separately than the numbers and thus we pack absolute values into a word and multiply the numbers as:

```
              1               2               5
       00000000000001  00000000000010  00000000000101
                                              10
X                                           001010
     ─────────────────────────────────────────────────
             10              20              50
  =    00000000001010  00000000010100  00000000110010         A
```

The sign bits are XORed together as

```
              +               +               -
        00000000000000  00000000000000  01000000000000
              +               +               +
XOR     00000000000000  00000000000000  00000000000000
        ─────────────────────────────────────────────
              +               +               -
=       00000000000000  00000000000000  01000000000000        B
```

Note that we use the second bit from left as the bit indicating the sign as if we use 2's complement representation we need the leftmost bit to take the possible carry. If we use the sign plus magnitude result we need OR A and B together as:

```
        00000000001010  00000000010100  00000000110010        A
OR      00000000000000  00000000000000  01000000000000        B
        ─────────────────────────────────────────────
=       00000000001010  00000000010100  01000000110010        Sign plus magnitude result
```

If we need to convert the result to 2's complement (in order for do addition) we need to shift B 12 bits (number of bits for each integer -2) to the left to get

```
        00000000000000  00000000000000  00000000000001        C
```

We then subtract C from B to get:

```
        00000000000000  00000000000000  01000000000000        B
-       00000000000000  00000000000000  00000000000001        C
        ─────────────────────────────────────────────
        00000000000000  00000000000000  00111111111111        D
```

We then AND A with D to extract negative numbers:

```
        00000000001010  00000000010100  00000000110010        A
AND     00000000000000  00000000000000  00111111111111        D
        ─────────────────────────────────────────────
        00000000000000  00000000000000  00000000110010        E
```

We then shift B to the left by 1 bit and subtract E to get 2's complement of E:

```
        00000000000000  00000000000000  10000000000000        B<<1
-       00000000000000  00000000000000  00000000110010        E
        ─────────────────────────────────────────────
        00000000000000  00000000000000  01111111001110        F, 2's complement of E
```

Now get nonnegative numbers by subtracting E from A:

```
    00000000001010 00000000010100 00000000110010        A
  - 00000000000000 00000000000000 00000000110010        E
    ─────────────────────────────────────────────
    00000000001010 00000000010100 00000000000000        G
```

Now OR F and G:

```
     00000000000000 00000000000000 01111111001110        F
  OR 00000000001010 00000000010100 00000000000000        G
     ─────────────────────────────────────────────
     00000000001010 00000000010100 01111111001110        2's complement result
```

Using the same technique we can get (2, 4, 6)x15, add (0, -3, 7) and then see the result of testing (10, 15) against all lines.

## 4. Locating a Point

For a query point $q=(x_0, y_0)$, we convert its coordinate real values to integers by cutting them at bit b (the same bit used to cut the points in the preprocessing stage) to convert them to a B+2 bit integers $x_1$ and $y_1$. We then test $(x_1, y_1)$ against all edges of all triangles. Thus we can decide for each triangle, whether $(x_1, y_1)$ is within the triangle, outside the triangle or on the edge of a triangle. Note that these triangle's edges' functions are $ax+by+c=0$ where a, b, c are converted integers.

These multiple integer addition, subtraction, multiplication is computed in constant time by packing them to a word as shown in the Section 3. These operations are used in previous research papers of other researchers. It is not clear who was the first to use these operations.

These operations are sufficient for us to determine the triangle that contain the integer coordinate of the query point in constant time.

## 5. Error Analysis

We showed that by packing integers into a word we can do point location in constant time. But when we convert real values to integers we introduce errors. These errors may make the judge of point location inaccurate. Here we discuss the error involved and show how to correct the errors.

Let the equation of an line be $ax+by+c=0$. Let the converted fraction values of a, b, c by $a_1$,

$b_1$, $c_1$. Let the query point location be $(x_0, y_0)$ and the converted fraction value be $(x_1, y_1)$. Let the error for converting a real number to a fraction number be err (this can be determined as the bit for cutting real numbers is known). Then the error for judging point $x_0$, $y_0$ against line $ax+by+c$ is:

$|ax_0+by_0+c-a_1x_1-b_1y_1-c_1|$
$< |ax_0+by_0+c-(a-err)(x_0-err)-(b-err)(y_0-err)+err|$
$< |a*err+x_0*err+b*err+y_0*err+2err^2+err|$

Let the maximum value of the coordinates be Max. Then we can let $err<1/Max^2$ to control the error. This can be done by let $b=b*2^{-2B}$. That is: instead of cutting B bits we cut 3B bits.

We can use this control strategy: if query point q is located within triangle T or on a edge of T (the converted fraction values) then we compare the real value of the point against the real values of the edges of T.

# 6.Main Theorem

**Main Theorem:** An n-vertex planar subdivision can be preprocessed in $O(n\sqrt{\log n})$ time and stored in $O(n)$ space to support constant time point location.

# 7.Conclusions

We studied the preprocess and point location for planar subdivision, We showed that n points subdivision can be preprocessed in $O(n\sqrt{\log n})$ time and stored in $O(n)$ space to support constant time point location. The $O(n)$ space is needed to store the real number values of the points to be compared with the query point, for if we do not need to compare the real number values of the points with the query point then the storage can be reduced to constant as we can pack all converted integer values in one word.